\pdfoutput=1
\documentclass[12pt,a4paper]{article}
\usepackage{ifthen} 
\newboolean{pdflatex}
\setboolean{pdflatex}{true} 

\newboolean{articletitles}
\setboolean{articletitles}{true} 

\newboolean{uprightparticles}
\setboolean{uprightparticles}{false} 

\newboolean{inbibliography}
\setboolean{inbibliography}{true} 

\usepackage{microtype}
\usepackage{lineno}  
\usepackage{xspace} 
\usepackage{caption} 
\usepackage[utf8]{inputenc}

\usepackage{graphicx}  
\usepackage{color}
\usepackage{colortbl}
\graphicspath{{./figs/}} 
\usepackage{xcolor}
\usepackage[most]{tcolorbox}
\usepackage[percent]{overpic}
\usepackage{amsmath} 
\usepackage{amssymb}
\usepackage{amsfonts}
\usepackage{upgreek} 

\DeclareUnicodeCharacter{2217}{ }

\newcommand*\patchAmsMathEnvironmentForLineno[1]{%
\expandafter\let\csname old#1\expandafter\endcsname\csname #1\endcsname
\expandafter\let\csname oldend#1\expandafter\endcsname\csname
end#1\endcsname
 \renewenvironment{#1}%
   {\linenomath\csname old#1\endcsname}%
   {\csname oldend#1\endcsname\endlinenomath}%
}
\newcommand*\patchBothAmsMathEnvironmentsForLineno[1]{%
  \patchAmsMathEnvironmentForLineno{#1}%
  \patchAmsMathEnvironmentForLineno{#1*}%
}
\AtBeginDocument{%
\patchBothAmsMathEnvironmentsForLineno{equation}%
\patchBothAmsMathEnvironmentsForLineno{align}%
\patchBothAmsMathEnvironmentsForLineno{flalign}%
\patchBothAmsMathEnvironmentsForLineno{alignat}%
\patchBothAmsMathEnvironmentsForLineno{gather}%
\patchBothAmsMathEnvironmentsForLineno{multline}%
\patchBothAmsMathEnvironmentsForLineno{eqnarray}%
}

\usepackage{hyperref}    
\usepackage[all]{hypcap} 

\usepackage{xspace} 
\usepackage{upgreek}

\def\lhcb {\mbox{LHCb}\xspace}

\def\babar  {\mbox{BaBar}\xspace}

\def\MagUp {\mbox{\em Mag\kern -0.05em Up}\xspace}

\ifthenelse{\boolean{uprightparticles}}%
{

 \def\Ppi         {\ensuremath{\uppi}\xspace}

 \def\PDelta      {\ensuremath{\Delta}\xspace}                 
 \def\PXi      {\ensuremath{\Xi}\xspace}                 
 \def\PLambda      {\ensuremath{\Lambda}\xspace}                 
 \def\PSigma      {\ensuremath{\Sigma}\xspace}                 
 \def\POmega      {\ensuremath{\Omega}\xspace}                 
 \def\PUpsilon      {\ensuremath{\Upsilon}\xspace}                 
 
 \def\PB      {\ensuremath{\mathrm{B}}\xspace}                 
                  
 \def\PD      {\ensuremath{\mathrm{D}}\xspace}

 \def\PK      {\ensuremath{\mathrm{K}}\xspace}

 \def\Pi      {\ensuremath{\mathrm{i}}\xspace}

 \def\Ps      {\ensuremath{\mathrm{s}}\xspace}

}
{

 \def\Ppi         {\ensuremath{\pi}\xspace}

 \mathchardef\PDelta="7101
 \mathchardef\PXi="7104
 \mathchardef\PLambda="7103
 \mathchardef\PSigma="7106
 \mathchardef\POmega="710A
 \mathchardef\PUpsilon="7107
                  
 \def\PB      {\ensuremath{B}\xspace}                 
                  
 \def\PD      {\ensuremath{D}\xspace}

 \def\PK      {\ensuremath{K}\xspace}

 \def\Pi      {\ensuremath{i}\xspace}

 \def\Ps      {\ensuremath{s}\xspace}

}

\makeatletter
\ifcase \@ptsize \relax
  \newcommand{\miniscule}{\@setfontsize\miniscule{4}{5}}
\or
  \newcommand{\miniscule}{\@setfontsize\miniscule{5}{6}}
\or
  \newcommand{\miniscule}{\@setfontsize\miniscule{5}{6}}
\fi
\makeatother

\DeclareRobustCommand{\optbar}[1]{\shortstack{{\miniscule (\rule[.5ex]{1.25em}{.18mm})}
  \\ [-.7ex] $#1$}}

\def\squark    {{\ensuremath{\Ps}}\xspace}

\def\pion   {{\ensuremath{\Ppi}}\xspace}
\def\piz    {{\ensuremath{\pion^0}}\xspace}

\def\kaon    {{\ensuremath{\PK}}\xspace}
  \def\Kbar    {{\kern 0.2em\overline{\kern -0.2em \PK}{}}\xspace}

\def\KorKbar    {\kern 0.18em\optbar{\kern -0.18em K}{}\xspace}

\def\Kp      {{\ensuremath{\kaon^+}}\xspace}
\def\Km      {{\ensuremath{\kaon^-}}\xspace}

\def\KS      {{\ensuremath{\kaon^0_{\mathrm{ \scriptscriptstyle S}}}}\xspace}

  \def\Dbar    {{\kern 0.2em\overline{\kern -0.2em \PD}{}}\xspace}
\def\D       {{\ensuremath{\PD}}\xspace}

\def\DorDbar    {\kern 0.18em\optbar{\kern -0.18em D}{}\xspace}
\def\Dz      {{\ensuremath{\D^0}}\xspace}
\def\Dzb     {{\ensuremath{\Dbar{}^0}}\xspace}
\def\Dp      {{\ensuremath{\D^+}}\xspace}

\def\Dstar   {{\ensuremath{\D^*}}\xspace}

\def\Dstarp  {{\ensuremath{\D^{*+}}}\xspace}

\def\Ds      {{\ensuremath{\D^+_\squark}}\xspace}

\def\B       {{\ensuremath{\PB}}\xspace}
\def\Bbar    {{\ensuremath{\kern 0.18em\overline{\kern -0.18em \PB}{}}}\xspace}

\def\BorBbar    {\kern 0.18em\optbar{\kern -0.18em B}{}\xspace}
\def\Bz      {{\ensuremath{\B^0}}\xspace}

\def\Bu      {{\ensuremath{\B^+}}\xspace}

\def\Bp      {{\ensuremath{\Bu}}\xspace}

  \def\Y#1S{\ensuremath{\PUpsilon{(#1S)}}\xspace}

\def\Lbar        {{\ensuremath{\kern 0.1em\overline{\kern -0.1em\PLambda}}}\xspace}
\def\LorLbar    {\kern 0.18em\optbar{\kern -0.18em \PLambda}{}\xspace}

\def\to                 {\ensuremath{\rightarrow}\xspace}

\def\AT#1     {\ensuremath{A_{\mathrm{T}}^{#1}}\xspace}           

\def\C#1      {\ensuremath{\mathcal{C}_{#1}}\xspace}                       
\def\Cp#1     {\ensuremath{\mathcal{C}_{#1}^{'}}\xspace}                    
\def\Ceff#1   {\ensuremath{\mathcal{C}_{#1}^{\mathrm{(eff)}}}\xspace}        
\def\Cpeff#1  {\ensuremath{\mathcal{C}_{#1}^{'\mathrm{(eff)}}}\xspace}       
\def\Ope#1    {\ensuremath{\mathcal{O}_{#1}}\xspace}                       
\def\Opep#1   {\ensuremath{\mathcal{O}_{#1}^{'}}\xspace}                    

\newcommand{\tev}{\ensuremath{\mathrm{\,Te\kern -0.1em V}}\xspace}
\newcommand{\gev}{\ensuremath{\mathrm{\,Ge\kern -0.1em V}}\xspace}
\newcommand{\mev}{\ensuremath{\mathrm{\,Me\kern -0.1em V}}\xspace}
\newcommand{\kev}{\ensuremath{\mathrm{\,ke\kern -0.1em V}}\xspace}
\newcommand{\ev}{\ensuremath{\mathrm{\,e\kern -0.1em V}}\xspace}
\newcommand{\gevc}{\ensuremath{{\mathrm{\,Ge\kern -0.1em V\!/}c}}\xspace}
\newcommand{\mevc}{\ensuremath{{\mathrm{\,Me\kern -0.1em V\!/}c}}\xspace}
\newcommand{\gevcc}{\ensuremath{{\mathrm{\,Ge\kern -0.1em V\!/}c^2}}\xspace}
\newcommand{\gevgevcccc}{\ensuremath{{\mathrm{\,Ge\kern -0.1em V^2\!/}c^4}}\xspace}
\newcommand{\mevcc}{\ensuremath{{\mathrm{\,Me\kern -0.1em V\!/}c^2}}\xspace}

\newcommand{\stat}{\ensuremath{\mathrm{\,(stat)}}\xspace}
\newcommand{\syst}{\ensuremath{\mathrm{\,(syst)}}\xspace}

\def\gsim{{~\raise.15em\hbox{$>$}\kern-.85em
          \lower.35em\hbox{$\sim$}~}\xspace}
\def\lsim{{~\raise.15em\hbox{$<$}\kern-.85em
          \lower.35em\hbox{$\sim$}~}\xspace}

\def\tell1  {TELL1\xspace}
\def\ukl1   {UKL1\xspace}

\usepackage{cite} 
\usepackage{mciteplus}

\usepackage{color} 
\graphicspath{{./}} 

\def\DsJ     {\ensuremath{\D_{sJ}^+}\xspace}
\def\dstarks   {\ensuremath{\Dstarp \KS}\xspace}

\def\Dstwo {\ensuremath{D^*_{s2}(2573)^{+}}\xspace}

\def\Dsa {\ensuremath{D^*_{s1}(2700)}\xspace}
\def\Dsja {\ensuremath{D^*_{sJ}(2700)}\xspace}

\def\Dsjb {\ensuremath{D_{sJ}^{*}(2860)}\xspace}
\def\Dsjan {\ensuremath{D^*_{sJ}(2700)}\xspace}
\def\Dsjbn {\ensuremath{D_{sJ}^{*}(2860)}\xspace}

\def\Dsju {\ensuremath{D_{sJ}(3040)}\xspace}

\def\calB         {{\ensuremath{\cal B}\xspace}}

\begin{document}


\renewcommand{\thefootnote}{\fnsymbol{footnote}}
\setcounter{footnote}{1}
\begin{titlepage}
\pagenumbering{roman}

\vspace*{2.0cm}

{\normalfont\bfseries\boldmath\large
\begin{center}
Experimental status of excited $D_s^+$ mesons 
\end{center}
}

\vspace*{1.0cm}

\begin{center}
  Antimo Palano, \\
  INFN Sezione di Bari, \\
  Via Orabona 4, 70125 Bari, Italy\\
  {\it antimo.palano@ba.infn.it}\\
  September 15, 2020
\end{center}
\begin{abstract}
  \noindent
The experimental status of the excited $D_s^+$ mesons is reviewed with particular emphasis on the most recent findings related to the $D^*_{s1}(2860)$ and $D^*_{s3}(2860)$ resonances. It is shown that the list of experimental results associated by the Particle Data Group to the observation of these states does not describe properly the experimental data.
\end{abstract}
\vspace*{2.0cm}

\end{titlepage}


\renewcommand{\thefootnote}{\arabic{footnote}}
\setcounter{footnote}{0}

\pagestyle{plain} 
\setcounter{page}{1}
\pagenumbering{arabic}

\section{Introduction}
\label{sec:Introduction}
\begin{figure}[ht]
  \begin{center}
    \includegraphics[width=0.9\linewidth]{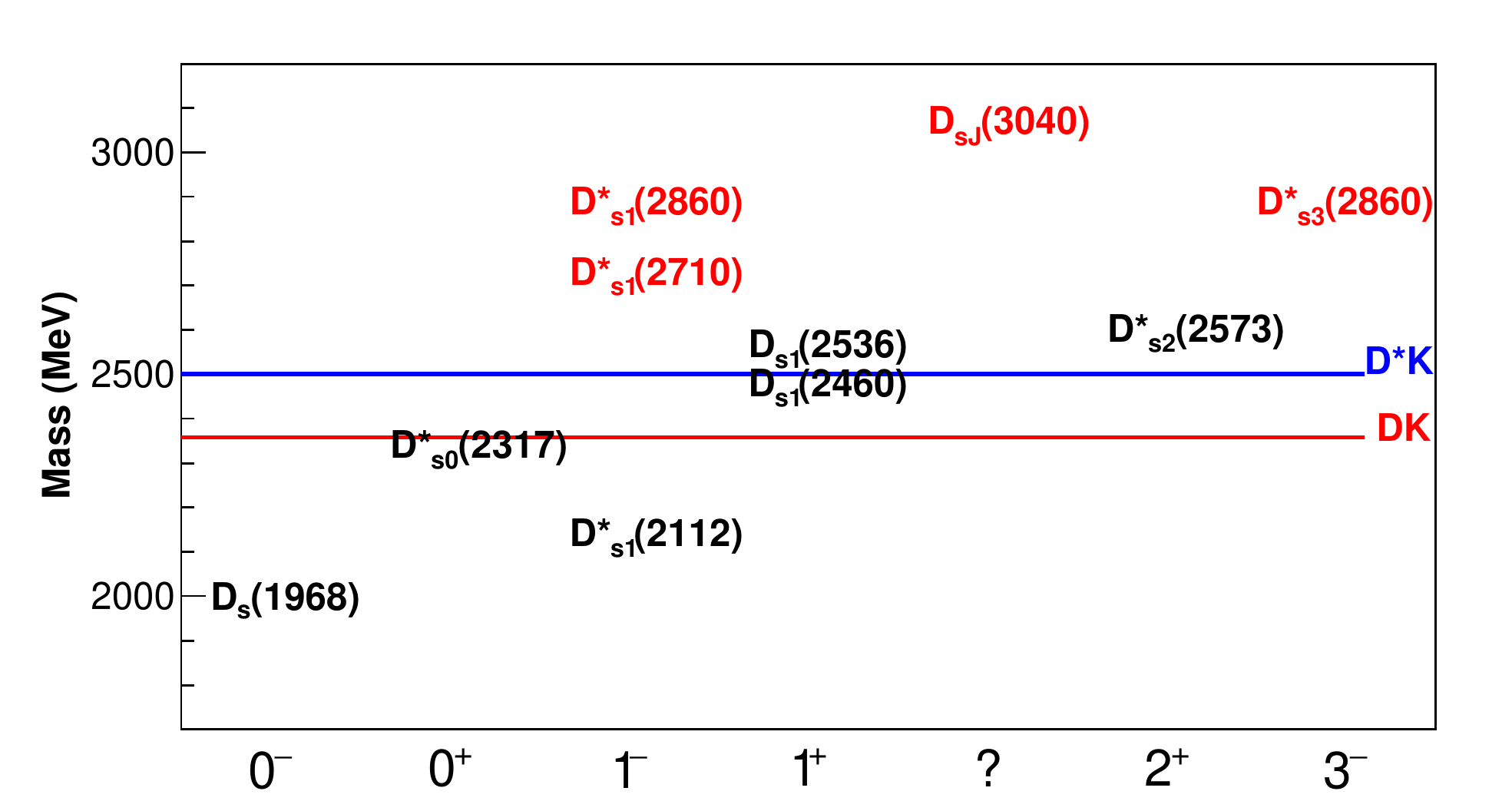}
    \vspace*{-0.5cm}
  \end{center}
  \caption{
    \small Experimental status of the $D_{sJ}$ mesons. The symbol (?) on the horizontal axis labels a signal having $J^P$ consistent with Unnatural Parity assignment.
    }
  \label{fig:fig1}
\end{figure}

The discovery by the \babar \ collaboration of a narrow state $D_{s0}^*(2317)^+$ in the decay to $\Ds\piz$~\cite{Aubert:2003fg}, and the subsequent discovery of a second narrow particle, $D_{s1}(2460)^+$ in the decay to $D_s^{*+}\piz$~\cite{Besson:2003cp,Abe:2003jk,Aubert:2003pe}, raised
considerable interest in the spectroscopy of heavy mesons.\footnote{The inclusion of charge-conjugate processes is implied, unless stated otherwise.}
These discoveries
were a surprise because quark model calculations based on heavy quark effective theory (HQET)~\cite{Isgur:1991wq}
predicted the masses of these resonances to be above the $DK$ and $D^*K$ thresholds, respectively.
Consequently their widths
were expected to be broad, as for the corresponding $J^P=0^+$ and $J^P=1^+$ resonances in the $D_J$ spectrum.
For a recent update of quark model status of the $D_{sJ}$ mesons spectroscopy see~\cite{Godfrey:2015dva}.
Figure~\ref{fig:fig1} summarizes the $D_{sJ}$ resonances listed in the Particle Data Book (PDG)~\cite{Zyla:2020zbs}.

\section{The observation of \Dsja and \Dsjb}

\begin{figure}[ht]
  \begin{center}
    \includegraphics[width=0.7\linewidth]{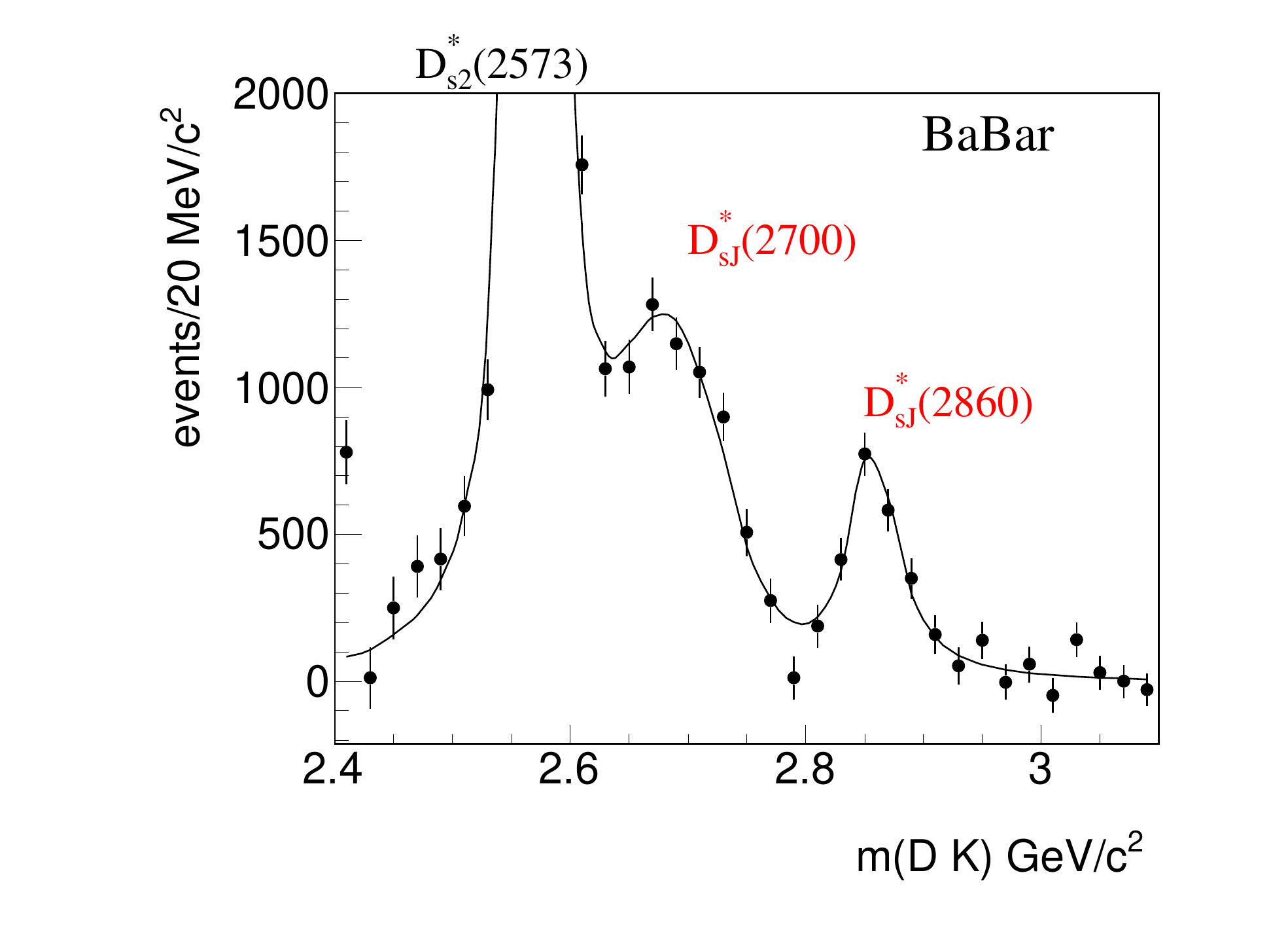}
    \vspace*{-0.5cm}
  \end{center}
  \caption{
    \small Total background subtracted $D K$ mass spectrum summed over the $D^0 \Kp$ and $D^+ \KS$ from BaBar~\cite{Aubert:2006mh}. Above the large $\D^*_{s2}(2573)$ signal, two natural parity states are observed for the first time.
    }
  \label{fig:fig2}
\end{figure}

The \DsJ mesons are expected to decay into the $D K$ and $D^*K$ final states if they are above threshold.
The BaBar collaboration has first explored the $D K$ mass spectrum~\cite{Aubert:2006mh} in inclusive reactions $e^+ e^- \to DK X$ observing two states, \Dsja and \Dsjb. The total, background subtracted mass spectrum, summed over the $D^0 \Kp$ and $D^+ \KS$ decay modes, is shown in fig.~\ref{fig:fig2}.

No spin-parity analysis in two-body decays is usually possible in inclusive reactions, however the decay of these states into two pseudoscalar mesons establishes a natural parity assignment.
States having $P=(-1)^J$ and therefore $J^P=0^+,1^-,2^+,...$ are referred as
natural parity states (NP) and are labeled as $D^*_s$, while unnatural parity (UP) indicates the series $J^P=0^-,1^+,2^-,...$ and states belonging to this series are labeled as $D_s$.

Tables~\ref{tab:tab1} and ~\ref{tab:tab2} summarize the properties of these states as measured by different experiments.
\begin{table}
\caption{Results from the fits to the  $DK$ and $\Dstar K$ mass spectra in different experiments for the \Dsjan meson. Resonance parameters are expressed in \mev.
When two uncertainties are presented, the first is statistical and the second systematic. Reference~\cite{Aubert:2009ah}, assuming the same resonances being observed decaying to $DK$ and $D^*K$, performed simultaneous fits to the $DK$ and $D^*K$ mass spectra. Therefore ref.~\cite{Aubert:2009ah} supersedes ref.~\cite{Aubert:2006mh}}.
\label{tab:tab1}
\centering
\resizebox{0.85\textwidth}{!}{
  \begin{tabular}{lcccc}
\hline
Data & $J^P$  & $m(\Dsjan)$ &  $\Gamma(\Dsjan)$ & Ref.\cr
\hline
$DK$, Inclusive, BaBar & NP & $2688 \pm 4  \pm 3$ & $112 \pm 7 \pm 36$ & \cite{Aubert:2006mh}\cr
$DK$, \Bp Dalitz, Belle & $1^-$ & $2708 \pm 9 ^{+11}_{-10}$ & $108 \pm 23 ^{+36}_{-31}$ & \cite{Brodzicka:2007aa}\cr
$DK$, \Bz Dalitz, BaBar &  $1^-$ & $2694 \pm 8 ^{+13}_{-3}$ & $145 \pm 24 ^{+22}_{-14}$ & \cite{Lees:2014abp}\cr
$DK$, \Bp Dalitz, BaBar &  $1^-$ &  $2707 \pm 8 \pm 8$ & $113 \pm 21 ^{+20}_{-16}$ & \cite{Lees:2014abp}\cr
$D^*K$, Inclusive, BaBar & NP & $2700 \pm 2^{+12}_{-7}$ & $149 \pm 7^{+39}_{-52}$ & \cite{Aubert:2009ah}\cr
$DK$, Inclusive, LHCb & NP & $2709.2 \pm 1.9 \pm 4.5$ & $115.8 \pm 7.3 \pm 12.1$ & \cite{Aaij:2012pc}\cr
$D^*K$, Inclusive, LHCb & NP & $2732.3 \pm 4.3 \pm 5.8$ & $136 \pm 19 \pm 24$ & \cite{Aaij:2016utb}\cr
\hline
\end{tabular}
}
\end{table}

\begin{table}
\caption{Results from the fits to the  $DK$ and $\Dstar K$ mass spectra in different experiments for the \Dsjbn meson. Resonance parameters are expressed in \mev.
When two uncertainties are presented, the first is statistical and the second systematic. Reference~\cite{Aubert:2009ah}, assuming the same resonances being observed decaying to $DK$ and $D^*K$, performed simultaneous fits to the $DK$ and $D^*K$ mass spectra. Therefore ref.~\cite{Aubert:2009ah} supersedes ref.~\cite{Aubert:2006mh}}.
\label{tab:tab2}
\centering
\resizebox{0.95\textwidth}{!}{
  \begin{tabular}{lcccc}
\hline
Data & $J^P$  & $m(\Dsjbn)$ & $\Gamma(\Dsjbn)$ & Ref.\cr
\hline
$DK$, Inclusive, BaBar & NP & $2856.6 \pm 1.5 \pm 5.0$ &$47 \pm 7 \pm 10$ & \cite{Aubert:2006mh}\cr
$D^*K$, Inclusive, BaBar & NP & $2862 \pm 2^{+5}_{-2}$ & $48 \pm 3 \pm 6$ & \cite{Aubert:2009ah}\cr
$DK$, Inclusive, LHCb & NP & $2866.1 \pm 1.0 \pm 6.3$ & $69.9 \pm 3.2 \pm 6.6$ & \cite{Aaij:2012pc}\cr
$D^*K$, Inclusive, LHCb & NP & $2867.1 \pm 4.3 \pm 1.9$ & $50 \pm 11 \pm 13$ & \cite{Aaij:2016utb} \cr 
\hline
\end{tabular}
}
\end{table}

\begin{figure}[ht]
\centering
 \begin{overpic}[clip,width=0.40\linewidth]{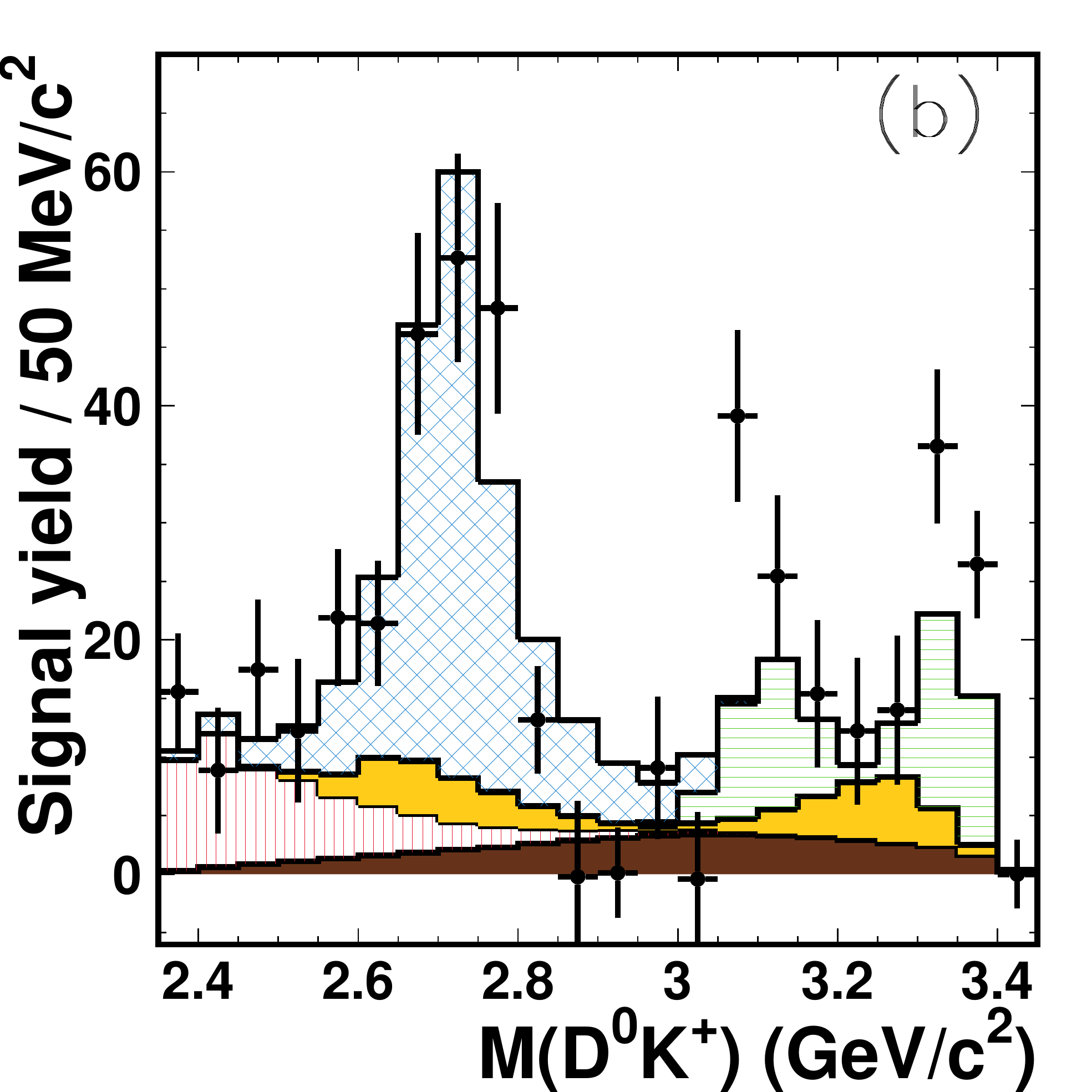}
  \put(50,74){\small\color{red}${\bf D^*_{s1}(2700)}$}
  \setlength{\fboxrule}{7pt}
\put(79,88){\fcolorbox{white}{white!60!white}{\null}}
\end{overpic}
    \includegraphics[width=0.55\linewidth]{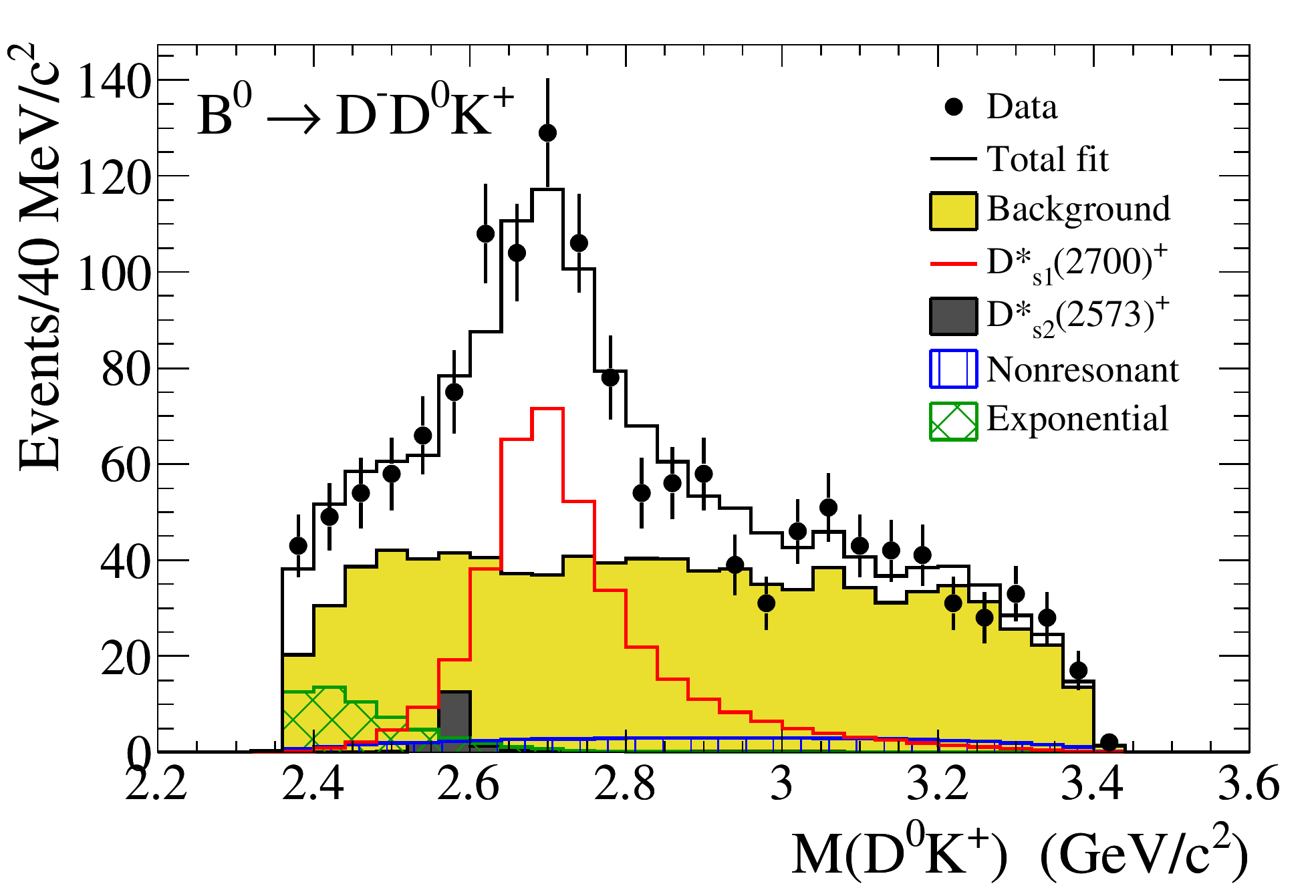}
  \caption{
    \small $D^0 K^+$ mass projection from the Dalitz plot analysis of $B^+ \to \Dz \Dzb \Kp$ from Belle~\cite{Brodzicka:2007aa} (left) and BaBar~\cite{Lees:2014abp} (right).
    }
  \label{fig:fig3}
\end{figure}

The \Dsja resonance was also observed by the Belle collaboration~\cite{Brodzicka:2007aa} and later, using $B^+$ and $B^0$ decays, by the BaBar collaboration~\cite{Lees:2014abp} in Dalitz plot analyses of $B$ decays to $\D\Dbar\kaon$. A full Dalitz plot analysis allows to obtain precise information on the spin-parity of the resonances contributing to the $B$ decay.
Both collaborations obtain a spin-parity assignment $J^P=1^-$ for this state, and therefore this state it is now labeled as \Dsa.
The $D^0 K^+$ mass projections from Belle and BaBar experiments are shown in fig.~\ref{fig:fig3} together with the fits results.

\begin{figure}[ht]
  \centering
   \begin{overpic}[clip,width=0.90\linewidth]{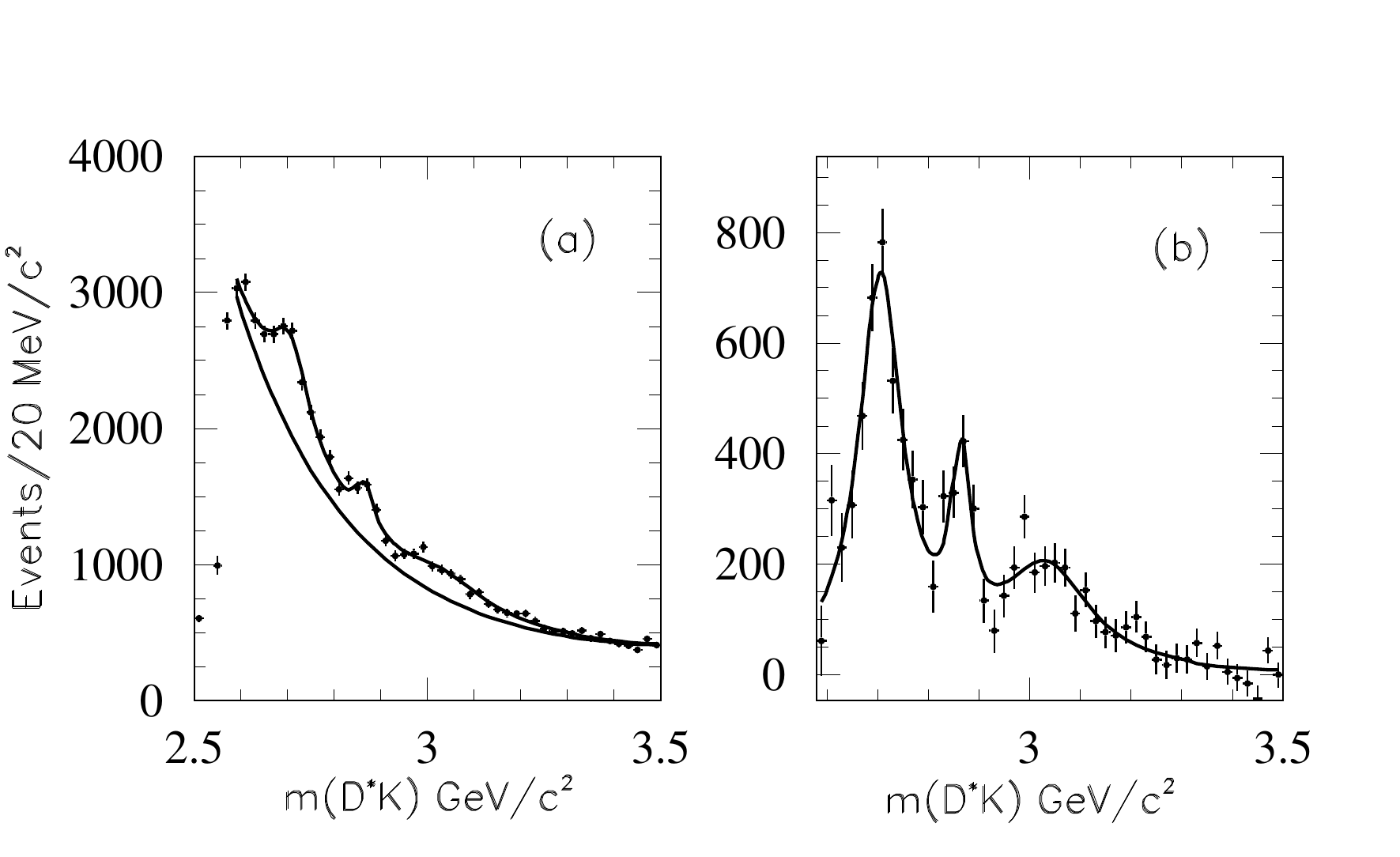}
  \put(65,44){\small\color{red}${\bf D^*_{s1}(2700)}$}
  \put(67,34){\small\color{blue}${\bf D^*_{sJ}(2860)}$}
  \put(75,24){\small\color{black}${\bf D_{sJ}(3040)}$}
  \end{overpic}
    \vspace*{-0.5cm}
  \caption{
    \small (a) Total $D^*K$ mass spectrum from BaBar summed over five different decay modes~\cite{Aubert:2009ah}. Above the $D_{s1}(2536)$ (not shown), three structures can be seen. (b) Background subtracted $D^*K$ mass spectrum.
    }
  \label{fig:fig4}
\end{figure}

Further information has been obtained by BaBar in the study of the inclusive production of the $D^*K$ mass spectrum, using five different final states. The total mass spectrum is shown in fig.~\ref{fig:fig4} and shows
signals \Dsjan and \Dsjbn, together with an additional new structure labeled as \Dsju. 

The observation of these resonances in the $D^*K$ final state allows to extract information on their spin-parity
assignment. In particular, it excludes $J^P=0^+$. In addition, the study of the helicity angle distribution $\theta_H$ allows
to have information on the naturality of these states. The helicity angle distribution, in fact, is expected to behave as $sin^2 \theta_H$ for natural parity and $1+hcos^2 \theta_H$ for unnatural parity, where the $h$ parameter depends on the state. A $cos^2 \theta_H$ distribution is expected for $J^P=0^-$. It is found that both \Dsjan and \Dsjbn angular distributions behave as $sin^2 \theta_H$ thus confirming a natural parity assignment. On the other hand the \Dsju structure is consistent with having an unnatural parity assignment.
The resulting fitted parameters are summarized in tables~\ref{tab:tab1} and ~\ref{tab:tab2}.

\begin{figure}[ht]
  \centering
    \setlength{\fboxrule}{5pt}
     \begin{overpic}[clip,width=0.47\linewidth]{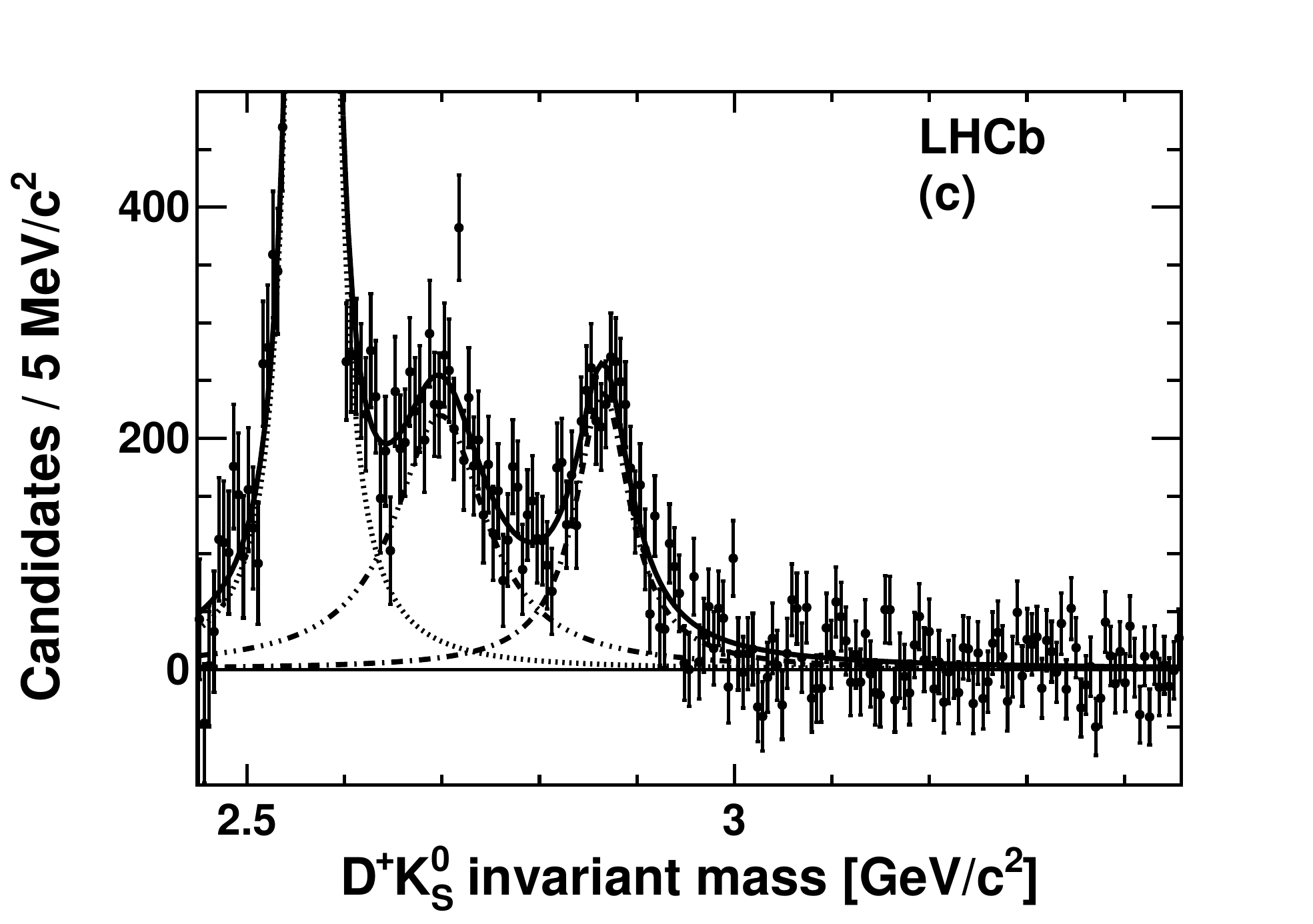}
     \setlength{\fboxrule}{5pt}
\put(70,54){\fcolorbox{white}{white!30!white}{\null}}
  \put(35,57){\small\color{red}${\bf D^*_{s1}(2700)}$}
  \put(45,47){\small\color{blue}${\bf D^*_{sJ}(2860)}$}
  \end{overpic}
       \begin{overpic}[clip,width=0.47\linewidth]{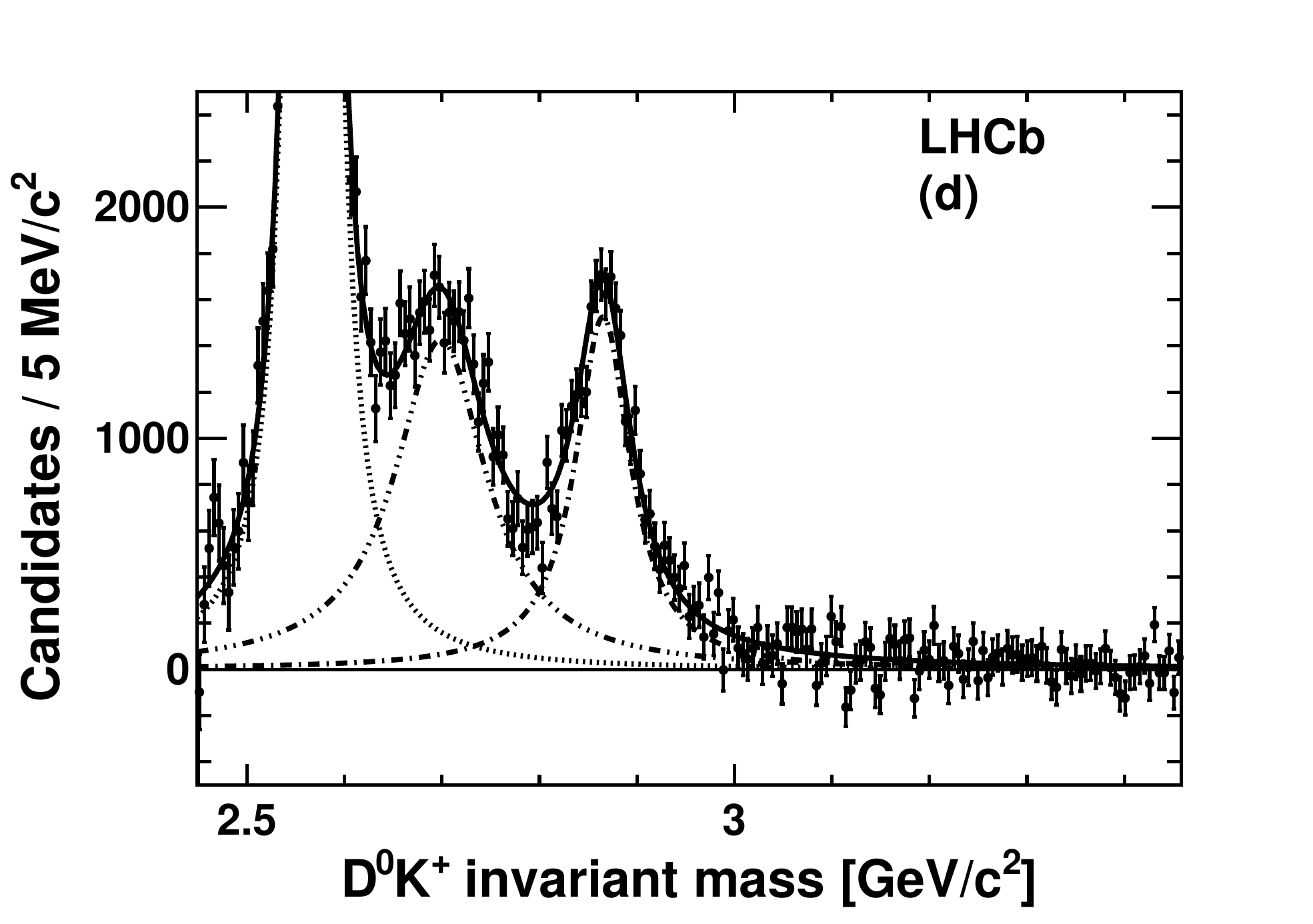}
\put(70,54){\fcolorbox{white}{white!30!white}{\null}}
  \put(33,54){\small\color{red}${\bf D^*_{s1}(2700)}$}
  \put(50,45){\small\color{blue}${\bf D^*_{sJ}(2860)}$}
  \end{overpic}
  \caption{
    \small Background subtracted $D^0 \Kp$ and $\Dp \KS$ mass spectra in inclusive $p p \to DK X$ from LHCb~\cite{Aaij:2012pc}.
    }
  \label{fig:fig5}
\end{figure}

The  \lhcb\ experiment has performed studies of the $DK$ and $D^*K$ final states in the inclusive process, $pp \to D^{(*)}K X$~\cite{Aaij:2012pc},~\cite{Aaij:2016utb}.
The \Dsa and \Dsjb are observed with large statistical significance and
their properties are found to be in agreement with previous measurements.
Figure~\ref{fig:fig5} shows the $DK$ mass spectra and fitted parameters are given in Tables~\ref{tab:tab1} and ~\ref{tab:tab2}.

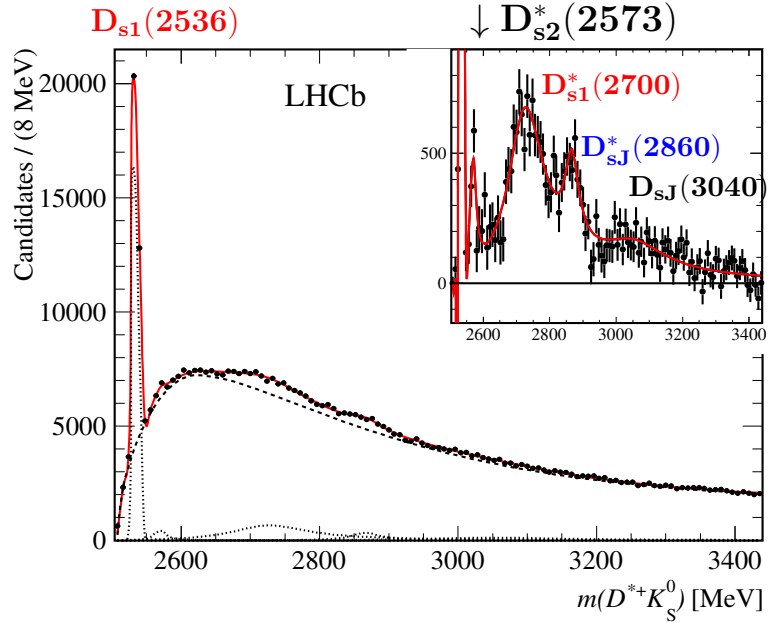
\begin{figure}[ht]
  \centering
     \begin{overpic}[clip,width=0.65\linewidth]{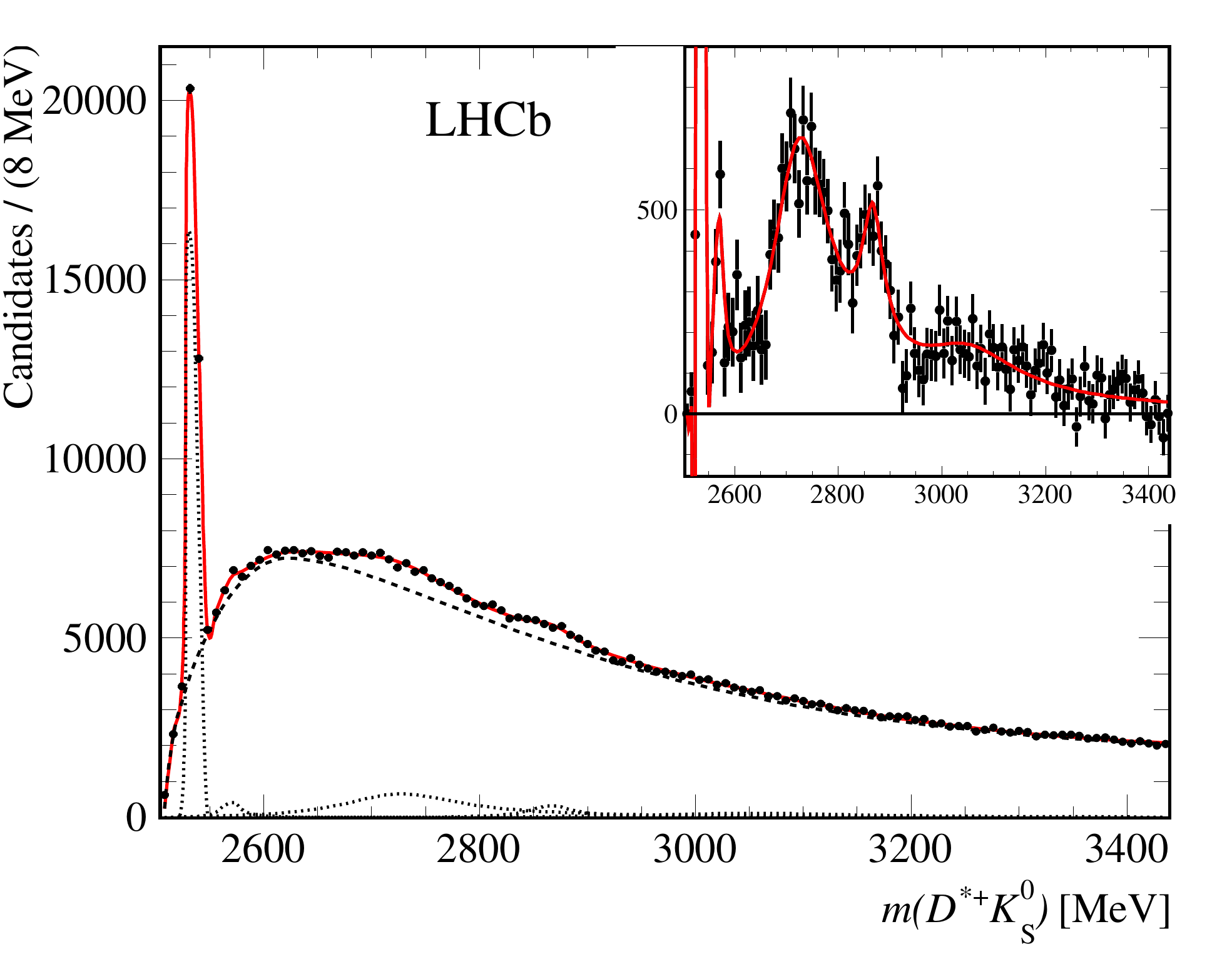}
   \put(10,76){\color{red}${\bf D_{s1}(2536)}$}
  \put(58,76){\large\color{black}$\downarrow{\bf D^*_{s2}(2573)}$}    
  \put(67,68){\small\color{red}${\bf D^*_{s1}(2700)}$}
  \put(72,60){\small\color{blue}${\bf D^*_{sJ}(2860)}$}
    \put(78,55){\small\color{black}${\bf D_{sJ}(3040)}$}
  \end{overpic}
  \caption{
    \small $D^{*+} \KS$ mass spectrum in inclusive $p p \to D^*K X$ from LHCb~\cite{Aaij:2016utb}. The inset shows the mass spectrum after
    the fitted background has been subtracted.
    }
  \label{fig:fig6}
\end{figure}

The $D^*K$ mass spectrum has been studied by LHCb using three different decay modes~\cite{Aaij:2016utb}.\footnote{This analysis and relative reported measurements have been omitted by PDG.} Figure~\ref{fig:fig6} shows the $D^{*+}\KS$ mass spectrum, where the \Dsa and \Dsjb resonances are again observed with fitted parameters given in Tables~\ref{tab:tab1} and ~\ref{tab:tab2}. The distributions of the helicity angle $\theta_H$ confirms the BaBar findings and therefore the natural parity assignment for these states.

In this LHCb analysis the \Dstwo decay to \dstarks is also observed for the first time, at a significance of $6.9\,\sigma$, with a branching fraction relative to the $\Dp \KS$ decay mode of
\begin{equation}
\frac{\calB(\Dstwo \to \Dstarp \KS)}{\calB(\Dstwo \to \Dp \KS)} = 0.044 \pm 0.005\stat \pm 0.011\syst.
\end{equation}
 This measurement is in agreement with expectations from recent calculations of the charm and charm-strange mesons spectra~\cite{Godfrey:2015dva} which predict a value of 0.058 for this ratio.
 
 \clearpage

\subsection{Summary on inclusive reactions}

\begin{figure}[h]
  \begin{center}
    \includegraphics[width=0.70\linewidth]{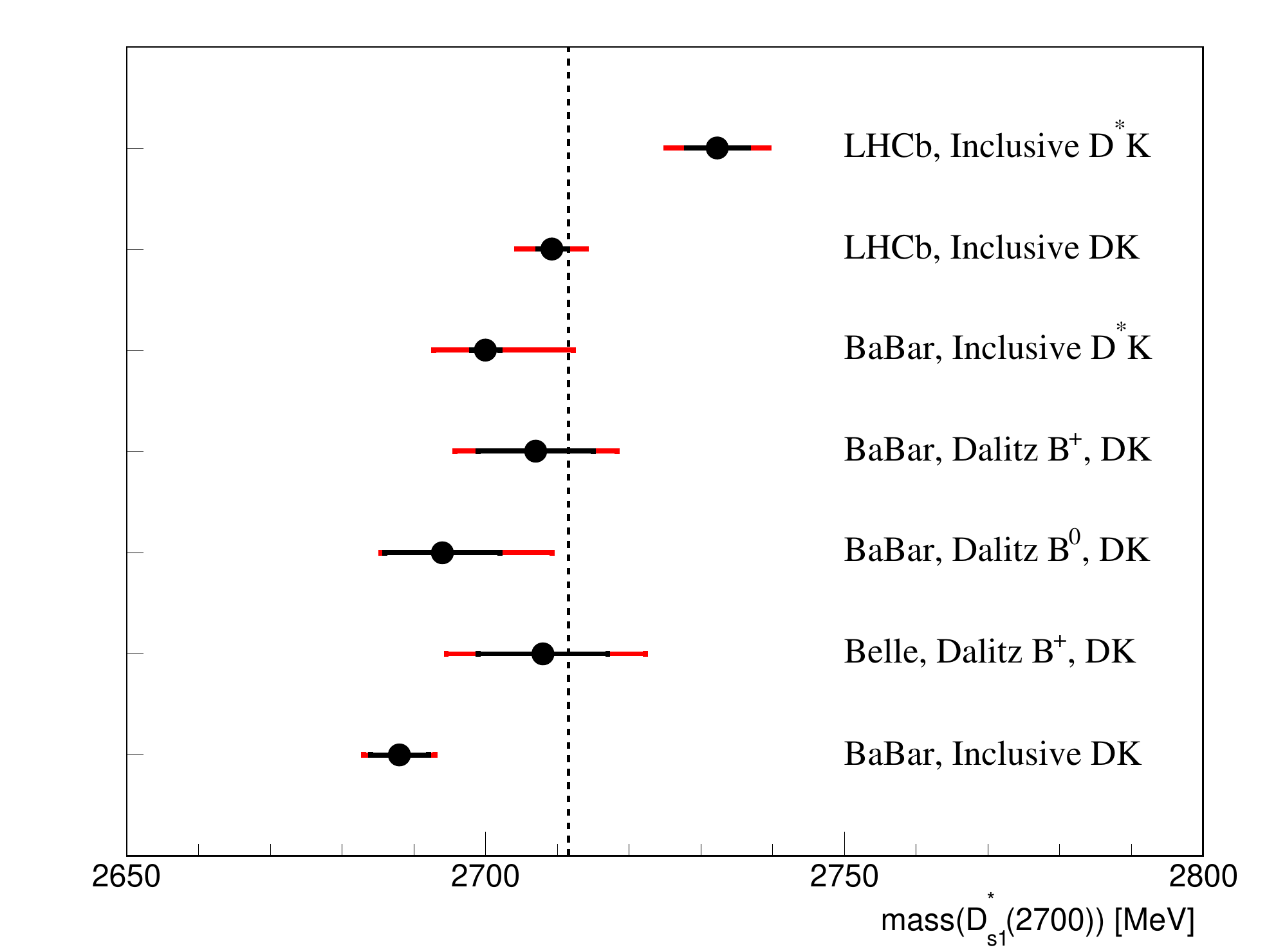}
   \includegraphics[width=0.70\linewidth]{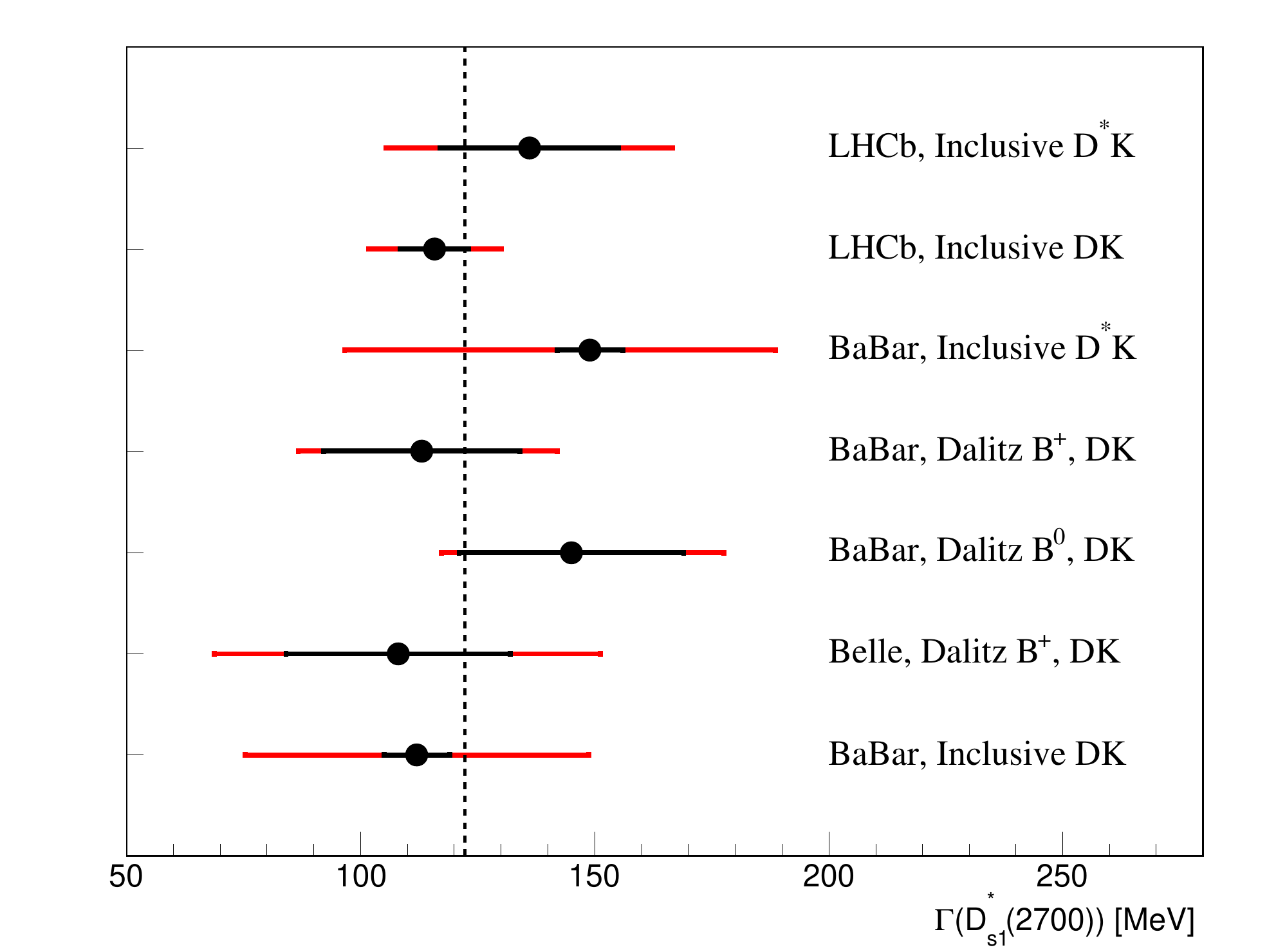} 
  \end{center}
  \caption{\small Summary of the measurement of the parameters of the $D^*_{s1}(2700)$ meson.
  The black line indicates the statistical uncertainties, the red line is obtained by adding in quadrature statistical and systematic uncertainties. The dashed line shows the results of a simplified weighted average where asymmetric systematic uncertainties have been averaged.}
  \label{fig:fig7}
\end{figure}

\begin{figure}[h]
  \begin{center}
    \includegraphics[width=0.70\linewidth]{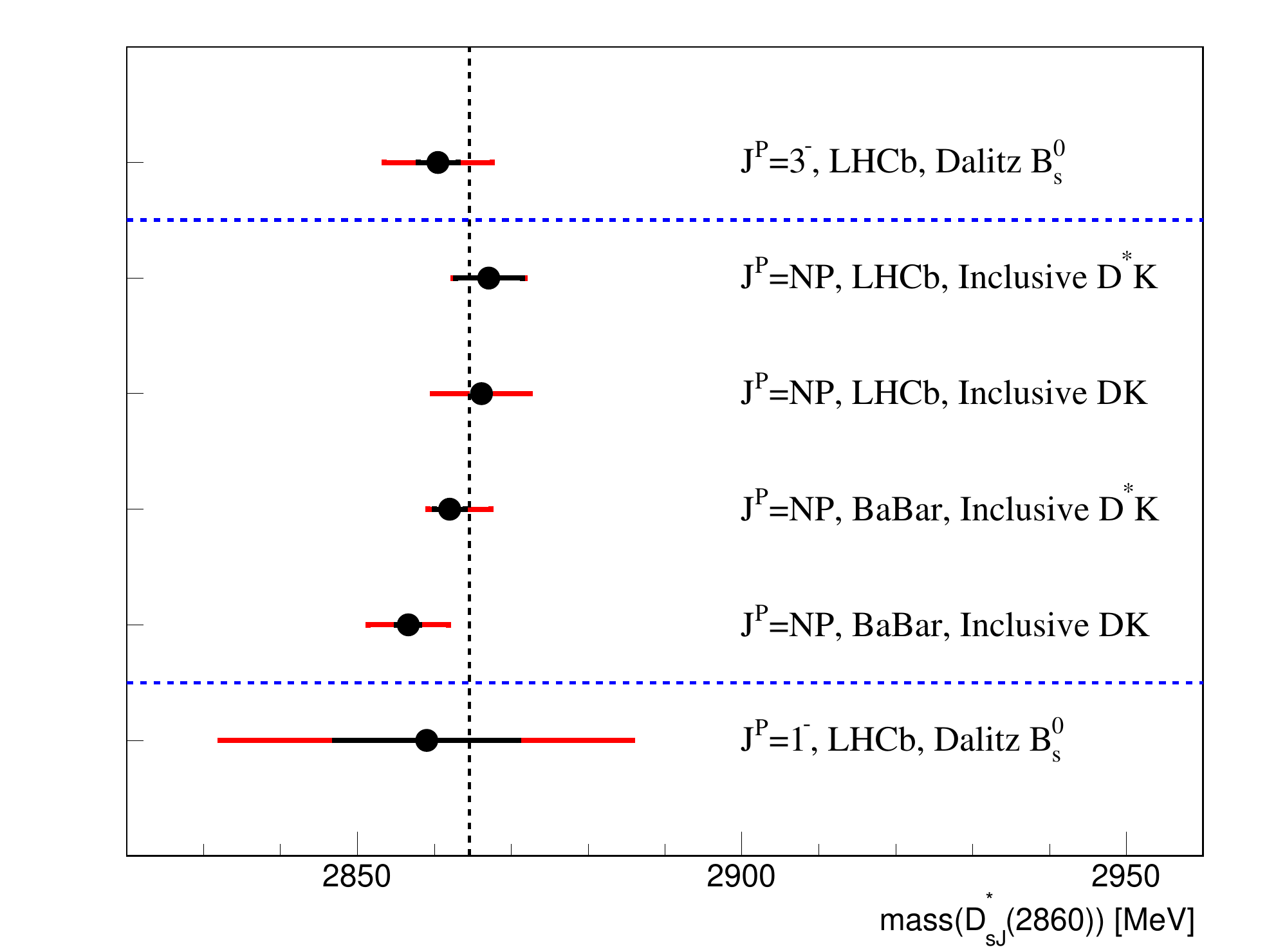}
   \includegraphics[width=0.70\linewidth]{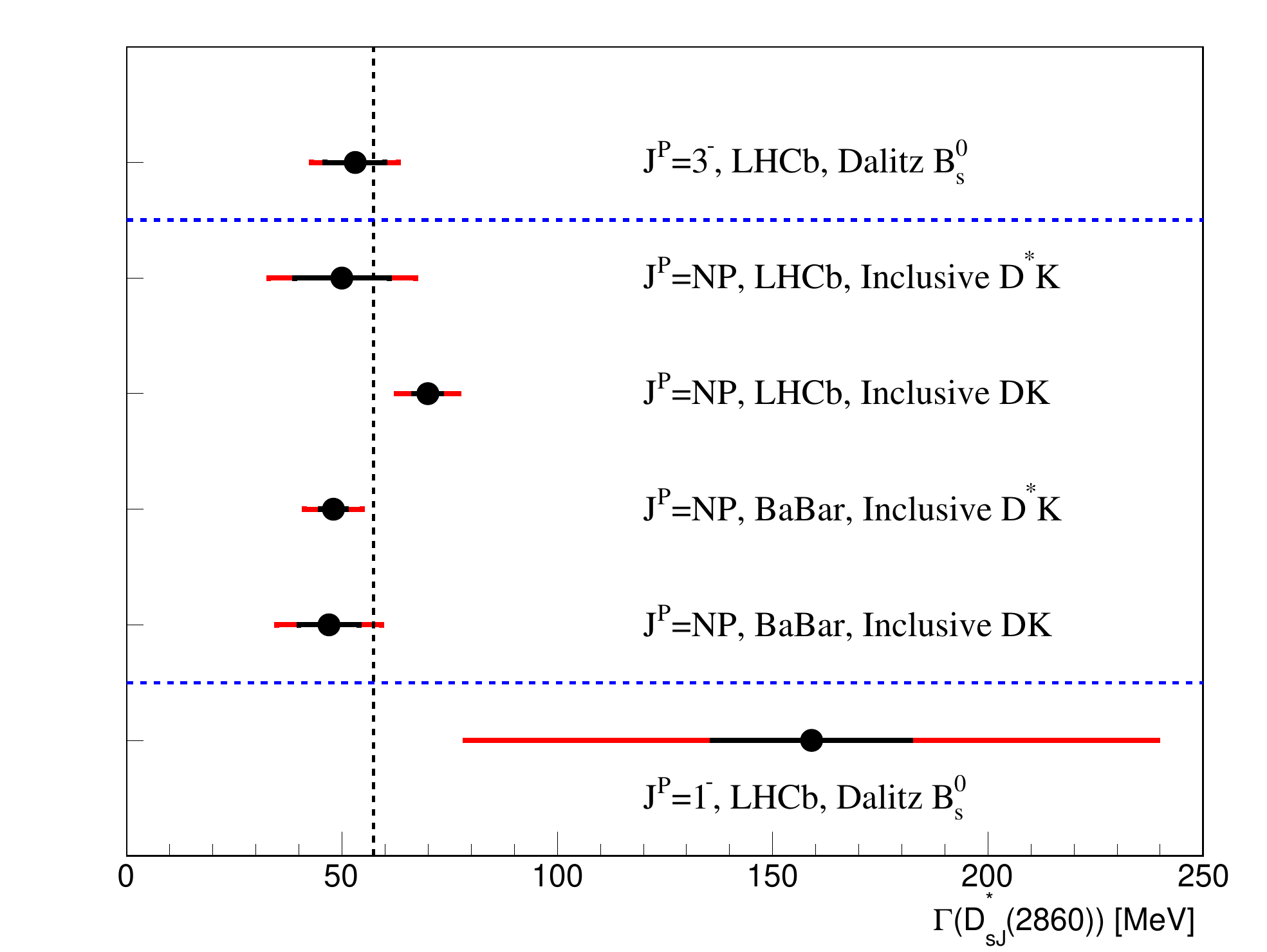} 
  \end{center}
  \caption{\small Summary of the measurement of the parameters of the $D^*_{sJ}(2860)$ meson.
  The black line indicates the statistical uncertainties, the red line is obtained by adding in quadrature statistical and systematic uncertainties. The dashed line shows the results of a simplified weighted average where asymmetric systematic uncertainties have been averaged. For comparison, the $D^*_{s1}(2860)$ and $D^*_{s3}(2860)$ parameters resulting from the $B_s^0$ Dalitz plot analysis are reported, but not included in the averaging procedure. When two systematic uncertainties are reported, they are added in quadrature.}
  \label{fig:fig8}
\end{figure}

The results of the measurements of the $D^*_{s1}(2700)$ parameters are summarized graphically in fig.~\ref{fig:fig7}.
The corresponding measurements of $D^*_{sJ}(2860)$ parameters are summarized graphically in fig.~\ref{fig:fig8}.

A simplified weighted average where asymmetric systematic uncertainties have been averaged gives the following parameters for the \Dsa meson:
\begin{equation}
m(\Dsa) = 2711.5 \pm 3.3 \ MeV, \qquad \Gamma=122 \pm 10 \ MeV
\end{equation}

and

\begin{equation}
m(\Dsjb) = 2864.6 \pm 2.8 \ MeV, \qquad \Gamma=57.4 \pm 4.8 \ MeV
\end{equation}

for the \Dsjb meson.

Several attempts have been made to identify these states within the quark model. In particular ref.~\cite{Colangelo:2012xi}
 and ref.~\cite{Godfrey:2015dva} point out that the spin-1 \Dsa can be identified with the $1^3D_1(c \bar s)$ state, while
 a spin-3 assignment is proposed for the \Dsjb which can be identified as the $1^3D_3(c \bar s)$ state.

\clearpage

\section{Results from the Dalitz plot analysis of $B_s^0 \rightarrow \Dzb K^- \pi^+$} 
 
\begin{figure} [h]
  \centering
\includegraphics[width=0.70\linewidth]{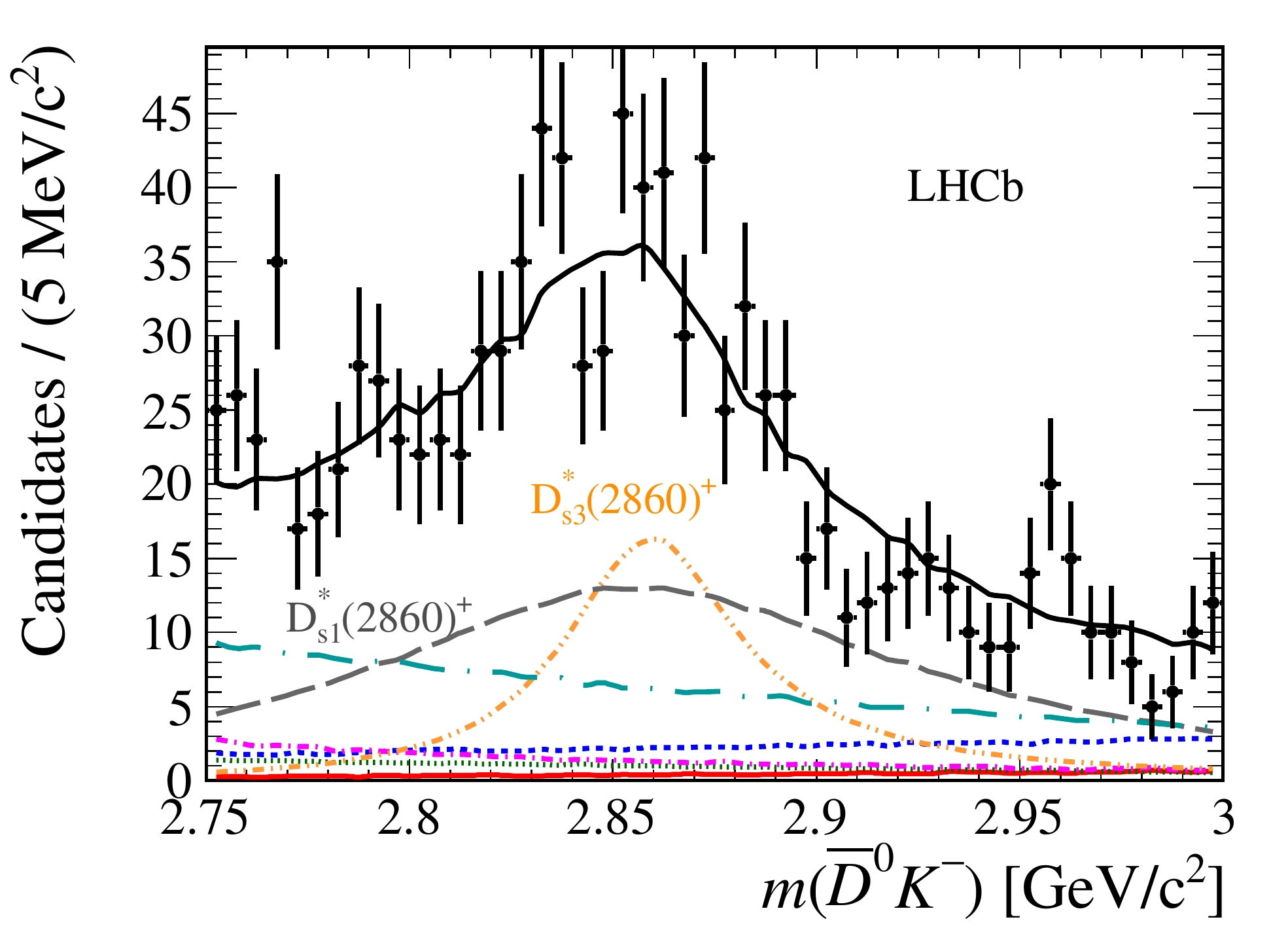}
  \caption{\small $\Dzb\Km$ mass projections from the Dalitz plot analysis of $B_s^0 \rightarrow \Dzb K^- \pi^+$ from LHCb experiment~\cite{Aaij:2014xza,Aaij:2014baa}.}
   \label{fig:fig9}
\end{figure}

LHCb has performed a Dalitz plot analysis of $B_s^0 \rightarrow \Dzb K^- \pi^+$ decays~\cite{Aaij:2014xza,Aaij:2014baa}.  Strong resonant production is observed both in the $K^- \pi^+$ and $\Dzb K^-$ systems. In particular, 
the $ \Dzb K^-$ mass 
spectrum shows a complex resonant structure in the $2860\,\mev$ mass region, shown in fig.~\ref{fig:fig9}. This is described by a superposition of a broad $J^P=1^-$ resonance and a narrow $J^P=3^-$ resonance with no evidence for the production of $D_{s1}^*(2700)$. The measured resonances parameters are:
\begin{equation}
  m(D^{*}_{s1}(2860))   =    2859    \pm 12   \pm 6    \pm 23  \mev,\ \Gamma(D^{*}_{s1}(2860)^-) = 159      \pm 23   \pm 27   \pm 72  \mev
  \end{equation}
  and
\begin{equation}  
m(D^{*}_{s3}(2860))       =  2860.5   \pm 2.6  \pm 2.5  \pm 6.0 \mev, \ \Gamma(D^{*}_{s3}(2860)^-)  =  53       \pm 7    \pm 4    \pm 6   \mev
\end{equation}
The fit fractions for the $D^{*}_{s1}(2860)$ and $D^{*}_{s3}(2860)$ components are $(5.0 \pm 1.2 \pm 0.7 \pm 3.3)\,\%$ and $(2.2 \pm 0.1 \pm 0.3 \pm 0.4)\,\%$, respectively, where the uncertainties are statistical, systematic and from Dalitz plot model variations.

\section{The PDG assignments}

The experimental status of the excited $D_s$ mesons is displayed in fig.~\ref{fig:fig7} and fig.~\ref{fig:fig8} and can be summarized as follows.

\begin{itemize}
\item{}
The \Dsa resonance, with $J^P=1^-$ decaying to $DK$ and $D^*K$ is well established both from inclusive and exclusive Dalitz plot analyses. Consistent measurements of its parameters are obtained from the different sources.

\item{}
A \Dsjb relatively narrow resonance is observed in inclusive reactions decaying to $DK$ and $D^*K$. Its spin-parity is natural, i.e. $J^P=1^-, 2^+, 3^-, ...$.

\item{}
A relatively narrow $D^{*}_{s3}(2860)$ resonance decaying to $DK$ is observed in the $B_s^0$ Dalitz plot analysis having parameters fully consistent with those of the \Dsjb resonance observed in inclusive reactions.

\item{}
A wide $D^{*}_{s1}(2860)$ resonance decaying to $DK$ is observed in the $B_s^0$ Dalitz plot analysis. The mass of this state is close to that of the \Dsjb observed in inclusive reactions but the width is about three times larger. The parameters of this state have large statistical and systematic uncertainties. Its contributing fraction to the $B_s^0$
decay is also affected by large model dependent uncertainties.

\end{itemize}

A natural summary of these experimental results would be to identify the $D^{*}_{s3}(2860)$ observed in $B_s^0$ decays with the \Dsjb resonance observed in inclusive reactions. The $D^{*}_{s1}(2860)$ resonance, given its rather undefined experimental status, and being seen only by one experiment, should wait for confirmation.

Starting from the 2016 PDG edition, although from the available experimental information the assignments $J^P=1^-, 2^+, 3^-, ...$ are possible, the \Dsjb resonance observed in inclusive reactions is assumed to have $J^P=1^-$ and associated to the  $D^{*}_{s1}(2860)$ state observed in $B_s^0$ decays. However, no average parameters are computed and all the results from inclusive reactions are listed as
``{\it ... We do not use the following data for averages, fits, limits, etc.}''

\clearpage

\section{Proposal}

On the basis of the actual experimental data it is proposed to update the PDG information related to the $D^{*}_{s3}(2860)$ as shown in Table~\ref{tab:tab3}. 
\begin{table} [h]
\caption{Proposed listing of the experimental results related to the $D^{*}_{s3}(2860)$ meson. Resonance parameters are expressed in \mev. Ref.~\cite{Aubert:2009ah} supersedes ref.~\cite{Aubert:2006mh}}.
\label{tab:tab3}
\centering
\resizebox{0.95\textwidth}{!}{
  \begin{tabular}{lcccc}
\hline
Data & $J^P$  & $m(D^{*}_{s3}(2860))$ & $\Gamma(D^{*}_{s3}(2860))$ & Ref.\cr
\hline
Averages & $3^-$ & $2862.2 \pm 2.6$ & $56.6 \pm  4.3$ & \cr
\hline
Dalitz $B_s^0$ & $3^-$ & $2860.5 \pm 2.6 \pm 2.5 \pm 6.0$ & $53  \pm 7    \pm 4    \pm 6$ & \cite{Aaij:2014xza,Aaij:2014baa} \cr
$D^*K$, Inclusive, LHCb & NP & $2867.1 \pm 4.3 \pm 1.9$ & $50 \pm 11 \pm 13$ & \cite{Aaij:2016utb} \cr
$DK$, Inclusive, LHCb & NP & $2866.1 \pm 1.0 \pm 6.3$ & $69.9 \pm 3.2 \pm 6.6$ & \cite{Aaij:2012pc}\cr
$D^*K$, Inclusive, BaBar & NP & $2862 \pm 2^{+5}_{-2}$ & $48 \pm 3 \pm 6$ & \cite{Aubert:2009ah}\cr
$DK$, Inclusive, BaBar & NP & $2856.6 \pm 1.5 \pm 5.0$ &$47 \pm 7 \pm 10$ & \cite{Aubert:2006mh}\cr
\hline
\end{tabular}
}
\end{table}

For what concerns the $D^*_{s1}(2860)$ observed in $B_s^0$ decays, its parameters are those in eq.(5). Its entry should contain the information ``{\it $J^P=1^-$ from angular analysis of \cite{Aaij:2014xza,Aaij:2014baa}}'' and ``{\it Separated from the spin-3 component $D^{*}_{s3}(2860)$ by a fit of the helicity angle of the $D^0K^-$ system, with a statistical significance of the spin-3 and spin-1 components in excess of 10$\sigma$}''.

\section{Summary}

The experimental status of the excited $D_s^+$ mesons is reviewed with particular emphasis on the most recent findings related to the $D^*_{s1}(2860)$ and $D^*_{s3}(2860)$ resonances. It is shown that the list of experimental results associated by the PDG to the observation of these states does not describe properly the experimental data.
It is proposed, on the basis of the present information on the fitted parameters and angular analyses,
that the $J^P=3^-$ $D^*_{s3}(2860)$
observed in $B_s^0$ and \Dsjb observed in inclusive reactions, are in fact the same state. On the other hand, the $J^P=1^-$ $D^*_{s1}(2860)$ only observed by one experiment should be waiting for confirmation.

\section{Acknowledgments}

I acknowledge Pietro Colangelo for useful comments.

\addcontentsline{toc}{section}{References}
\bibliographystyle{LHCb}

\bibliography{main}

\end{document}